# Business and Regulatory Responses to Artificial Intelligence: Dynamic Regulation, Innovation Ecosystems and the Strategic Management of Disruptive Technology[*]

Mark Fenwick, Erik P. M. Vermeulen and Marcelo Corrales Compagnucci

**Abstract** Identifying and then implementing an effective response to disruptive new AI technologies is enormously challenging for any business looking to integrate AI into their operations, as well as regulators looking to leverage AI-related innovation as a mechanism for achieving regional economic growth. These business and regulatory challenges are particularly significant given the broad reach of AI, as well as the multiple uncertainties surrounding such technologies and their future development and effects.

This chapter identifies two promising strategies for meeting the "AI challenge," focusing on the example of Fintech. First, "dynamic regulation," in the form of regulatory sandboxes and other regulatory approaches that aim to provide a space for responsible AI-related innovation. An empirical study provides preliminary evidence to suggest that jurisdictions that adopt a more "proactive" approach to Fintech regulation can attract greater investment. The second strategy relates to so-called "innovation ecosystems." It is argued that such ecosystems are most effective when they afford opportunities for creative partnerships between well-established corporations and AI-focused startups and that this aspect of a successful innovation ecosystem is often overlooked in the existing discussion.

The chapter suggests that these two strategies are interconnected, in that greater investment is an important element in both fostering and signaling a well-functioning innovation ecosystem and that a well-functioning ecosystem will, in turn, attract more funding. The resulting synergies between these strategies can, therefore, provide a jurisdiction with a competitive edge in becoming a regional hub for AI-related activity.

# 1 Introduction

Finding an appropriate response to disruptive new AI technologies is enormously challenging for any business looking to integrate AI into their operations or for regulators looking to leverage AI-related innovation as a means of achieving regional economic growth. These

---

business and regulatory challenges are particularly significant given the potential reach of AI, as well as the multiple uncertainties surrounding such technologies and their future development and effects.

This chapter begins with a brief overview of three key features of the "AI challenge" (Section 2). The three features identified are (i) the disruption of traditional business models triggered by AI-technologies (Section 2.1); (ii) the increase in AI-driven investment and the new opportunities and resulting disruption that this has triggered (Section 2.2); and (iii) the profound uncertainties that surround the possible future development and effects of AI-related technologies (Section 2.3). Each of these issues has significant implications for business and regulators.

Meeting the AI challenge is crucial for established corporations, startups and policymakers. The chapter identifies two promising strategies for regulating AI based on the experience of regulating previous disruptive new technologies. First, "dynamic regulation," in the form of so-called "regulatory sandboxes" and other proactive regulatory approaches, that aim to provide a space for responsible AI-related business innovation. An empirical study provides preliminary evidence to suggest that jurisdictions that adopt a more proactive approach to the regulation of Fintech do indeed seem to attract greater investment. Section 3 outlines some of the main features of such an approach.

The second regulatory strategy relates to so-called "innovation ecosystems." It is suggested that such ecosystems are particularly effective when they afford opportunities for more creative partnerships between established corporations and AI-focused startups and that this feature of ecosystems is often neglected in the existing discussion. The main features and potential benefits of such partnerships are outlined in Section 4.

The chapter argues that these two strategies are inter-connected, in that greater investment is an important element in fostering a well-functioning innovation ecosystem. The resulting synergies between these two strategies can, therefore, provide a jurisdiction with a competitive edge in an effort to become a regional hub for AI-related activities.

Although these regulatory strategies are not riskless (either for business or the state), they do, nevertheless, represent the best option for responding to the AI challenge. At least, they seem to be clearly preferable to the two obvious alternatives, namely strict ex ante control (which risks stifling innovation, investment, and growth, or – at least – prompting a "brain drain" or capital flight) or a socially-irresponsible de-regulation (which may result in harmful effects, particularly in the context of technologies whose effects are uncertain).

## 2 The "AI Challenge"

It is helpful to begin by distinguishing three important aspects of the "AI challenge," at least as it impacts upon business and government (regulators and other policy-makers).

## 2.1 AI-Technologies & the Disruption of Existing Business Models

When thinking about the business and regulatory challenges that are created by AI technologies, it is important to adopt a broad-based definition of AI. Such a definition encompasses all four main types of AI technology:

i. "Type 1" AI refers to purely "reactive" machines that specialize in one area or task. For instance, the drafting and review of loan agreements. More "famous" examples would be IBM's Deep Blue chess software or Google's Alpha Go algorithm for playing Go;
ii. "Type 2 AI" machines *possess just enough memory or "experience" to make proper decisions and execute appropriate actions in specific situations or contexts.* Self-driving cars, chatbots, or personal digital assistants are the most commonly cited examples;
iii. "Type 3" AI has the capacity to understand thoughts and emotions which affect human behavior. Softbank Robotic's "Pepper" can organize large amounts of data and information in order to have a "human-like" conversation;
iv. "Type 4 AI" is "artificial intelligence" as it is typically portrayed in Hollywood movies or TV shows (think HBO's Westworld). Machines using this type of AI are self-aware, super-intelligent, sentient and are presumed to possess something like consciousness.

The advantage of an expansive definition of this kind is that it highlights the urgency of AI-related developments and avoids the risks of complacency. If we define AI narrowly in terms of Type 4 AI – i.e., AI that is "more human than human" – we don't need to be overly concerned with the disruptive impact and potential effects of such technologies at the moment. After all, "singularity" – the moment when AI capacities surpass our own – seems to be some decades away. The advantage of a broader definition of AI is that it, therefore, allows us to appreciate the extent and diversity of the business challenges that are *already* created by AI and allows us to develop a portfolio of possible regulatory strategies appropriate for managing such technologies. Moreover, all aspects of business operations seem likely to be affected by at least one of these four types of AI, and such disruption is already occurring across multiple sectors of the economy.

Take the example of financial services. There are many tasks that are central to financial services that can already be better performed by machines. Type 1 AI can do certain things more effectively than a human (for instance, reviewing large numbers of standard form contracts). Even though Type 4 AI may be a long way off that doesn't mean that the financial services industry isn't already being disrupted by other, "simpler," forms of machine intelligence.

Fintech – broadly defined as the use of new technologies to make financial services, ranging from online lending to digital currencies, more efficient – can already be seen across a range of financial services and many of these new services involve some kind of AI, in the broad sense adopted here.[1] For example, p*eer-to-peer lending platforms that use algorithms and machine learning to assess the creditworthiness of borrowers. Or, "robo-advisors" that automate many aspects of personal finance and wealth management. Such intelligent machines can already help individuals manage their personal accounts, debts, assets and investments.*[2]

Or, in the context of healthcare and life sciences: artificial intelligence already takes the role of an experienced clinical assistant that can help doctors make faster and more reliable diagnoses. We already see AI applications in the areas of imaging and diagnostics, and oncology, for example.[3] More generally, machine learning has the potential to improve remote patient monitoring. AI algorithms are able to take information from electronic health records, prescriptions, insurance records and even wearable sensor devices to design a personalized treatment plan for patients. Finally, AI-related technologies accelerate the discovery and creation of new medicines and drugs. There is a broad consensus amongst insiders that healthcare is being transformed for the better as a result of AI. And the opportunities and potential are limitless.

Broadly defined, it is hard to imagine any existing business that isn't profoundly disrupted by AI. Artificial intelligence, machine learning, and deep learning are just the beginning of a revolution that will transform everyday life and how we interact with technology. And new AI-oriented start-ups looking to satiate this new demand are rapidly emerging, which brings us to the second aspect of the AI challenge.

## *2.2 AI-Driven Investment, Start-Ups & a New Market for Corporate Control*

The result of these technological and business developments is that AI is attracting record amounts of investors' money. Consider global venture capital investments in AI, which increased significantly between 2012 and 2016, both in terms of absolute amount invested and the number of deals (see Figure 1).

---

[1] For a general introduction to Fintech, see Arner et al. (2016a), (2016b); Haddad and Hornuff (2016); MIT Sloan School of Management (2016). For a discussion of Fintech and investment-related issues and trends, see Fenwick, McCahery and Vermeulen (2018).
[2] See, generally, Fenwick and Vermeulen (2017).
[3] For a general introduction, see JASON (2017).

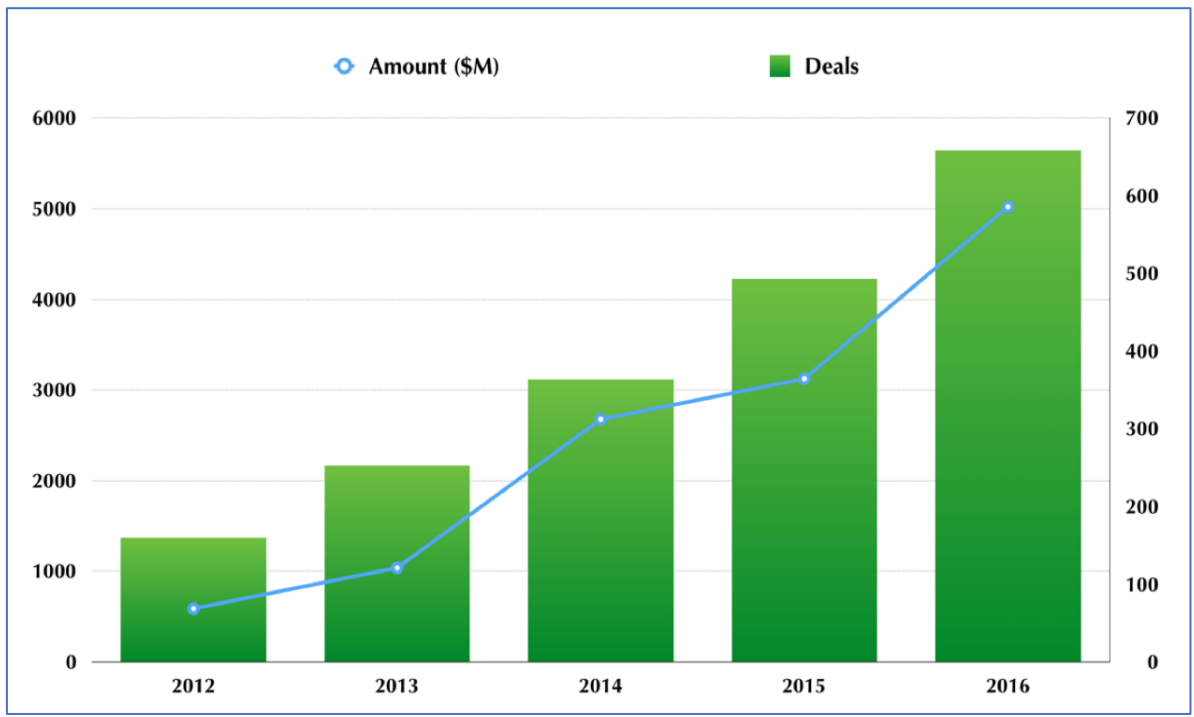

*Figure 1:* Global Venture Capital Investments in AI by Amount & No. of Deals (2012-16)[4]

Moreover, M&A activity involving AI companies has increased significantly over a similar five-year time-scale (see Figure 2). Typically, the acquired companies in such acquisitions are often Silicon Valley-based startups and the majority are from the U.S (see Figure 3). But that doesn't necessarily make it the U.S. the only AI center in the world. Even a high-level review of the available data suggests that other regions are active and that AI entrepreneurs can be found everywhere.

It seems clear that established corporations and investors value new companies that embrace these new technologies and gradually bring them to market. It is hardly surprising that have replaced the financial institutions and oil businesses as the largest companies in the world, at least according to their market capitalization.

And if we consider some of the world's largest companies – think Apple, Alphabet, Microsoft or Amazon – we can see that these companies view different types of Artificial Intelligence as a key business opportunity for the future. Siri (Apple), Google Assistant (Alphabet), Cortana (Microsoft) and Alexa (Amazon) are already able to assist you with more and more difficult tasks, and this trend is only set to continue.

---

[4] Source: CB Insights.

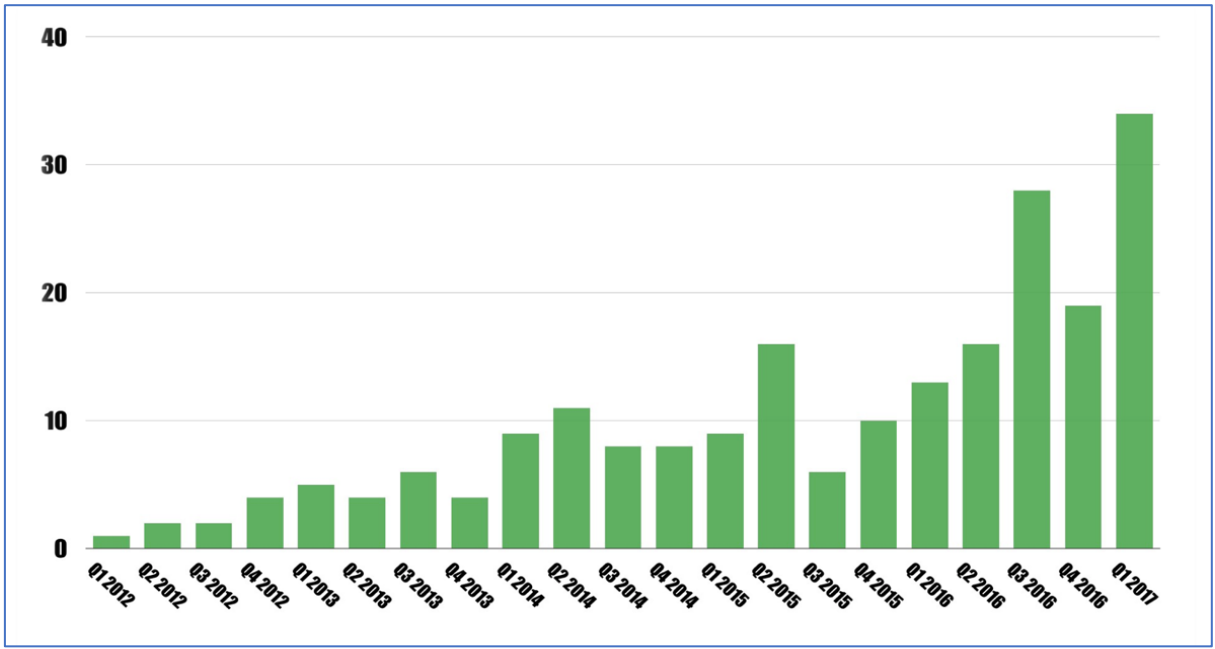

*Figure 2:* *M&A Activity Involving AI Companies by Number of Deals (2012-17)[5]*

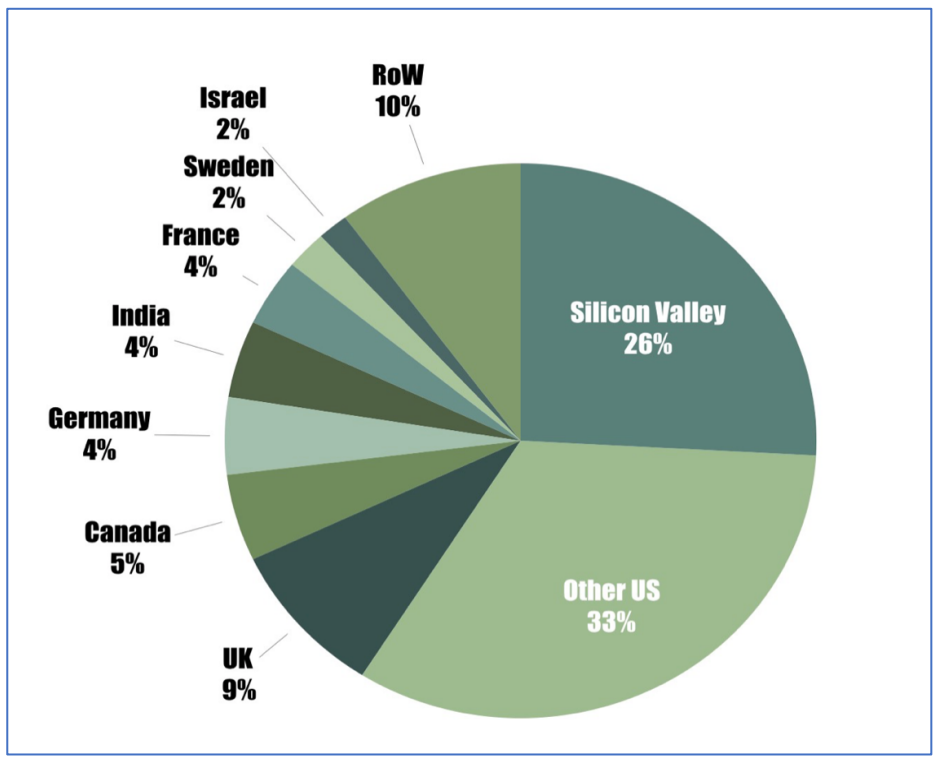

*Figure 3:* *The Location of Acquired AI Companies (2012-2017)[6]*

---

[5] Source: CB Insights.
[6] Source: Pitchbook.

## 2.3 AI Technologies & Radical Uncertainty

A final point regarding the challenge of AI technologies. As Types of AI develop more uncertainties will inevitably occur, particularly as we approach "singularity", and this highlights a transformation in the character of risk. A paradox of digital technologies is that they make our lives easier, but they also make the world harder – perhaps even impossible – to understand. The digital world is a world of risk – of identifiable and measurable dangers – but, more significantly, it is also a world of radical uncertainty.[7] Our relationship with new technology is often characterized by uncertainty, in the sense that "all we know is that there are many things that we do not know" about a technology and its effects.

The speed of technological development means that transformative change will come much sooner than expected. Big data and the near-endless amounts of information have undoubtedly transformed AI to unprecedented levels. Blockchain technology and smart contracts will merely continue and, very probably, accelerate the trend. The enormous increase in computational power, the breakthrough of "Internet of Things" applications and the further development of smart machines will only accelerate AI's development and global adoption. The acceleration of innovation will add to AI's ability to adapt to new situations and solve problems that currently seem to be impossible.

Any list of potential outcomes – positive or negative – created by new technologies is always going to be incomplete. As such, the digital world is a world where "reality" and "truth" regarding new technologies are uncertain, unsettled and constantly being contested.

In part, this is simply a function of the ever-quickening speed of technological change.[8] As soon as we believe that we have a clear understanding, a new development has already occurred that renders any existing understanding obsolete. But something else is also going on. "Understanding" of complex man-made systems is now increasingly "beyond" human comprehension.[9] For the first time in history, we live in a world where more and more technologies are simply beyond human comprehension.

So how can business and regulators meet this AI challenge? In this chapter, two potentially fruitful strategies are introduced and explored. First, new forms of regulation, notably regulatory sandboxes (Section 3), and, second, innovation ecosystems that foster partnerships between established firms and startups (Section 4).

# 3 "Responsive" / "Smart" / "Dynamic" Regulation

Recently, the regulation literature – particularly that branch of the discussion focusing on the regulation of new and disruptive technologies has focused on improving the ability of regulators to respond to changing industry practices (especially technology-driven changes) and the ability to improve relationships between regulators and regulated companies.[10] So-called dynamic regulation can respond to changing industry practices through feedback effects and enhanced information for regulation.

---

[7] See Beck (1992).
[8] See Bennett Moses (2011).
[9] See Arbesman (2016).

In this context, what is particularly important is that that within the framework of these new and more dynamic models, regulatory decisions should not be thought of as 'final events' (to be made for all-time and from which we "all move on"). Rather, we should think of regulatory choices as a form of "measured decision-making," i.e., regulatory choices are open-ended and highly contingent selections that are merely one stage or element in a longer narrative and not the "final word" on a particular issue. As such, regulators need to abandon a fixation on finality and legal certainty and embrace contingency, flexibility and an openness to the new. The justification for this new openness derives from the contingency of the technology-dominated environment in which regulators now must operate.

This shift in perspective affects how we regulate disruptive technologies. Rather than approaching decisions as "final events" (to be made for all-time and to which we all commit), Michel Callon has proposed the alternative notion of "measured action" (i.e., measured decision-making), where you do not decide an outcome, you take contingent measures that are based on inclusive processes involving both experts and the public.[11] Any regulatory "choice" ultimately remains open-ended, leaving space to incorporate new knowledge, discoveries, and claims. The need for finality, Callon argues, is usually overstated, more the product of expediency and habit than actual necessity.

Similarly, Gralf-Peter Caliess and Peer Zumbasen's concept of "rough consensus and running code" developed in the context of transnational business law also highlights a new contingency as a defining feature of contemporary "law"-making in transnational settings.[12]

According to this perspective, the antidote to the "Hobson's choice" of recklessness (i.e., an irresponsible deregulation or non-regulation) versus paralysis (i.e., an excess of regulation that stifles innovation) in the regulation of disruptive technologies is a willingness to remove the temporal horizon that has traditionally defined decision-making, while at the same time creating new and more dynamic mechanisms for consistent citizen involvement in the ongoing process of determining measured action.

Of course, this may very well be a noble goal, but the problem is how to operationalize such an approach in more concrete terms. An obvious solution to this regulatory dilemma might be to adopt some form of policy experimentation, i.e., testing different regulatory schemes and then comparing the results. But such experimentation poses a problem for regulators. Too often, regulators define "success" in negative terms, as in the avoidance of catastrophe:

> "One defining feature of the Strong Precautionary Principle is that it places a governmental entity in a role as a risk gatekeeper. Implicit in the Principle is the idea that there must be a 'decider' who will determine whether the proponent of the activity has met its burden of proof on safety. The preventive thrust of Strong

---

[10] See, e.g., Black, Hopper and Band (2007); Black (2009); Kaal (2013); Kaal (2014).
[11] Callon et al. (2009).
[12] Calliess and Zumbansen (2010).

> Precaution further implies that this review of risks should occur *before* the activity commences or the potentially risky product reaches the market."[13]

Avoiding grounds for criticism inevitably results in an overly cautious approach, often called the "precautionary principle." In this regard, the recent "rise" of the so-called "regulatory sandbox" is particularly interesting as a concrete approach to regulation that ensures a responsible regulatory framework that doesn't have a chilling effect on technological innovation.

### 3.1 Regulatory Sandboxes

In the financial industry, it has recently been suggested that such an engaged-approach with disruptive technology is best facilitated by the establishment of regulatory sandboxes.

The Financial Conduct Authority (FCA), the financial regulatory body in the United Kingdom, is widely accredited with first introducing this approach.[14] In April 2016, the FCA broke new ground by announcing the introduction of a regulatory sandbox which allows both start-up and established companies to "test" new ideas, products and business models in the area of Fintech.

The aim of the sandbox is to create: a "safe space" in which businesses can test innovative products, services, business models and delivery mechanisms without immediately incurring all the normal regulatory consequences of engaging in the activity in question.

The idea behind the sandbox is for the state regulator to approve a firm-specific, de-regulated space for the testing of innovative products and services without being forced to comply with the applicable set of existing rules and regulations. With the sandbox, the regulator aims to foster innovation by lowering regulatory barriers and costs for testing disruptive innovative technologies, while ensuring that consumers will not be negatively affected. The three key questions that were investigated by the FCA on the sandbox proposal concerned "regulatory barriers" (how and to what extent can they be lowered?), "safeguards" (what protector measures should be in place to ensure safety), and "legal framework" (what regulatory arrangement is mandated by EU law).

What is perhaps most interesting about the sandbox is that new ideas, products and services can be tested in a "live," "real world" environment. As such, firms are given the authorization to test their products or strategies without being subjected to existing regulatory requirements and the associated prohibitions or compliance costs. In order to create this environment, the FCA has defined a set of default parameters that can be altered on a case-by-case basis, depending on the needs of a particular firm. These parameters include:

i. *Duration.* As a default, the FCA considers three to six months to be an appropriate length of time to 'test' a particular innovation.

---
[13] Sachs (2010), p. 1298; see also Sunstein (2005).
[14] See Financial Conduct Authority (2015).

ii. *Scale.* The number of customers should be big enough to generate statistically relevant data and information on the product or service. This means that customers should be selected based on certain criteria that are appropriate for the product and service. Clearly, pre-agreed safeguards and protections should be in place to protect consumers.
iii. *Prior Disclosure.* Customers should be accurately informed about the test and any available compensation (if needed). Moreover, indicators, parameters and milestones that are used during the testing phase should be clearly set out from the outset.

What makes the regulatory sandbox model so attractive is that, insofar as technology has consequences that flow into everyday lives, such technology will be open to discussion, democratic supervision and control. In this way, public entitlement to participate in regulatory debates can help to create a renewed sense of legitimacy and confidence that justifies the regulation that is subsequently adopted.

It comes as no surprise that regulatory sandboxes are being adopted by other regulators, such as the Australian Securities and Investment Commission, Singapore's Monetary Authority and Abu Dhabi's Financial Services Regulatory Authority.

In discussions about regulatory sandboxes with other experts in banking and finance, arguments are often made suggesting that their deployment is nothing more than a strategy of a country to signal its openness to innovation and technology. In their view, "sandboxes" aren't really offering anything new. Regulators are usually able to exempt companies and technologies to comply with the applicable set of rules and regulations without referring them to a sandbox. The Australian "Fintech" exemption is an example.

Yet, these arguments seem to miss the main advantages of the "regulatory sandbox." The potential of regulatory sandboxes goes much further than a signal. Insofar as technology has consequences that flow into everyday lives, such technology will be open to discussion and democratic supervision and control. In this way, public entitlement to participate in regulatory debates can help to create a renewed sense of legitimacy that justifies the regulation.

What is even more important is that regulatory sandboxes offer opportunities to generate information and data relevant to the regulation of the new digital world. They allow the participants in the sandbox, i.e., regulators, incumbent companies, start-ups, investors, consumers, to learn about the new technologies (such as AI). In this way, they can create the necessary dialogue that helps us understand new technologies. They allow for collaboration and joint discovery. But perhaps most importantly, they create an opportunity to change the mind-set of incumbents operating in the financial service sector and allow them to embrace the new possibilities associated with artificial intelligence, machine learning and deep learning.

In an age of constant, complex and disruptive technological innovation, knowing prioritized *what, when,* and *how* to structure regulatory interventions has become much more difficult. Regulators can find themselves in a situation where they believe they must opt for either reckless action (regulation without sufficient facts) or paralysis (doing nothing). Inevitably in such a case, caution tends to be over risk. The precautionary principle becomes the default position. But such caution merely functions to reinforce the status quo and the result is that new technologies struggle to reach the market in a timely or efficient manner.

*3.2 An Empirical Test*

In order to explore empirically the effects of a more dynamic regulatory approach to disruptive new technologies, a small empirical study was conducted, focusing on the example of Fintech.

Broadly speaking, if we look around the world today we can distinguish between two broad categories of regulatory response – "reactive" and "proactive" – both of which comprise a number of sub-categories.

On the one hand, there are what we can characterize as reactive jurisdictions. This includes countries in which nothing is being done, i.e., there is currently no regulatory talk or action responding to Fintech. A second sub-group consists of those countries in which there is only partial or fragmented regulation of Fintech. Certain institutions, such as the Consumer Financial Protection Bureau in the United States, may offer certain safe harbor provisions for certain type of Fintech companies. Yet, there appears to be little willingness to genuinely embrace the technology and its regulatory implications, nor is there any comprehensive plan as to how Fintech can or should be regulated.

On the other hand, are those countries that take a more a proactive approach. In this group, we find those countries that make Fintech a strategic priority. In such countries, more regulatory attention paid to Fintech. Again, a number of sub-categories can be identified.

A first sub-group comprises countries in which such "attention" takes the form of consultation papers, White Papers, or conferences. Of course, there is a risk that such "talk" can slide into empty "lip service" aimed at projecting an image of regulatory action when, in reality, action is limited or non-existent.

A second sub-group of countries engage in what we might characterize as "regulatory guidance." Regulators issue guidelines or provide advice to Fintech start-ups and incumbents in order to help navigate them through the regulatory system. This does not necessarily entail changes in formal regulatory structures, but it does provide some support innovation. The initiative to issue a national charter for the supervision of Fintech companies by the U.S. Office of the Controller of the Currency is a recent example.

A final group of countries has embraced the possibilities of Fintech by creating a regulatory sandbox, as described above. We would characterize this approach as "regulatory experimentation." Regulators create a sandbox in which they facilitate and encourage a space to experiment. This allows the testing of new technology-driven services, under the supervision of regulators. This ensures that meaningful data can be gathered for the evaluation of risk in a safe environment. Such data can then facilitate "evidence-based regulatory reform." A key point about this last approach is that it is collaborative and dialogical, in the sense that regulators, incumbents and new service providers are engaged in an on-going dialogue about the most effective means to gather relevant information and to identify the most appropriate regulatory model.

In order to better understand, the effects, risks and opportunities associated with these regulatory choices, we conducted a simple empirical study of regulatory responses to Fintech in twelve jurisdictions.

In particular, we looked at first-time "venture capital" investments in Fintech companies. The intention was to see whether there was a meaningful connection between levels of investment and the regulatory choice reactive or proactive. When we look at the results of Year-on Year %-growth of first-time venture capital backed companies we get the following Figure 4. In many cases, this data confirms anecdotal evidence of a slow-down of interest in Fintech. But interestingly, in six of the twelve jurisdictions, there was an increase in investment activity in 2016.

The question this data raises is whether there are any signals as to a correlation between more proactive regulatory initiatives and increased activity in the Fintech sector? In those countries in which the response was reactive (red line), there seems to be some evidence of a slowdown. In contrast, in those countries with a more proactive response – particularly involving regulatory guidance (green "dashed" line) or regulatory experimentation (solid green line) – there is some evidence to suggest that this proactive approach makes the jurisdiction more attractive as a potential location for starting Fintech operations (see Figure 5).

The above analysis suggests in a preliminary way that the regulatory environment does affect the degree of investment and – perhaps as importantly – the willingness of companies to start operations in one jurisdiction, rather than another. In this respect, "regulation matters." This is not to underestimate difficulties of finding an appropriate regulatory regime:

> "One obstacle to this goal is that new technologies are often met with highly polarized debates over how to manage their development, use and regulation. Prominent examples include nuclear energy and genetically modified foods…These technologies are characterized by a rapid pace of development, a multitude of applications, manifestations and actors, pervasive uncertainties about risks, benefits and future directions, and demands for oversight ranging from potential health and environmental risks to broader social and ethical concerns. Given this complexity, no single regulatory agency, or even group of agencies, can regulate any of these emerging technologies effectively and comprehensively."[15]

It is precisely for this reason that the more dynamic and experimental approach associated with "sandboxes" seems so promising.

---

[15] Mandel et al. (2013), pp. 45 and 136; see, generally, Marchant and Wallach (2013).

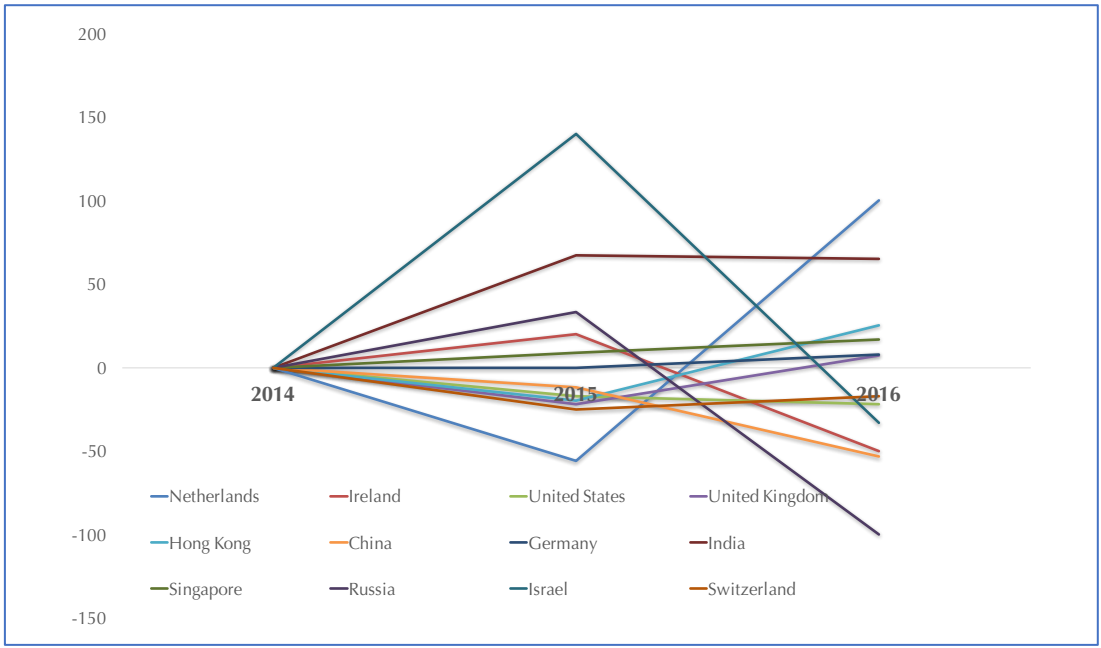

*Figure 4:* Year-on-Year % Growth of "First-time" Venture Capital-Backed Fintech Companies (by Country, 2014-16)[16]

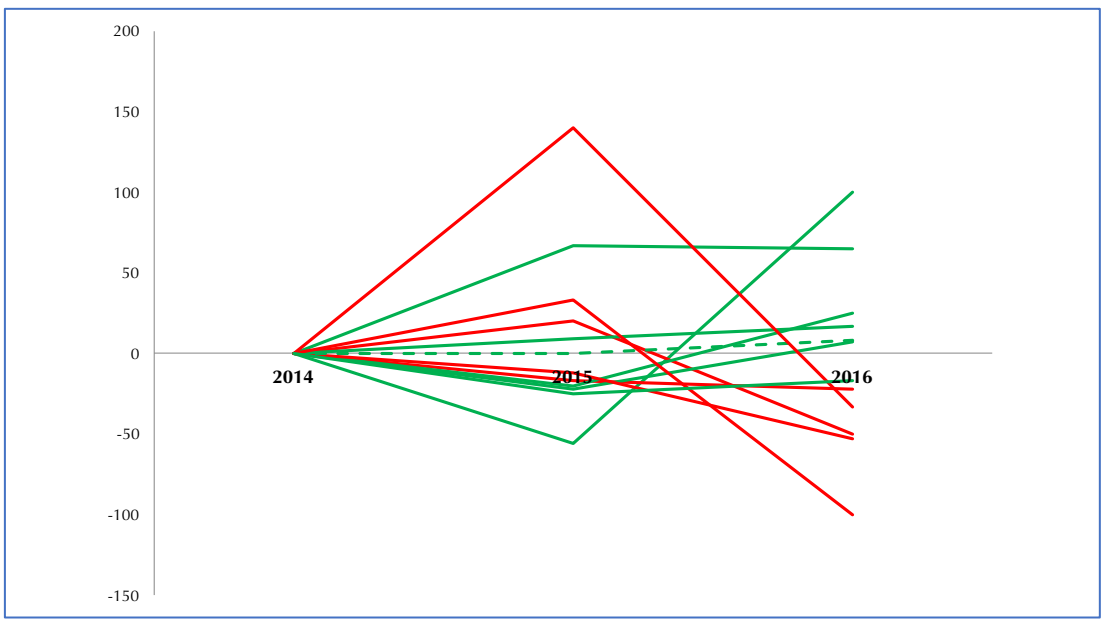

*Figure 5:* Year-on-Year % Growth of "First-time" Venture Capital-Backed Fintech Companies (by Regulatory Approach, 2014-16)[17]

## 4 Innovation "Ecosystems"

---

[16] Source: Pitchbook.
[17] Source: Pitchbook.

Recognizing the importance of more responsive forms of regulation in the context of disruptive technologies is only part of the story, however. We also need to acknowledge that there are other considerations that make a particular ecosystem attractive for Fintech or other AI-related industries. This suggestion points us towards discussion of "innovation systems" and why it is that particular regions become the focus of innovative activity.

## *4.1 Replicating Silicon Valley*

Research has consistently shown that over the last three decades Silicon Valley has been the place to go for anyone interested in setting up a new business, particularly a tech-related business.[18] It has consistently ranked as the best location for launching a new business with global aspirations. Silicon Valley attracts the most funding, it is the most connected and it offers the most opportunities for both innovators and entrepreneurs. Silicon Valley has represented the best bet for anyone with serious aspirations of creating a global business in high growth sectors of the economy.

As a result of this success, policymakers have been drawn to the idea of recreating the success of Silicon Valley in other parts of the world.[19] Initially, this discussion focused on strategies for promoting investment. Generally, this involved two primary types of public intervention into the venture capital market, namely (i) by regulation or de-regulation, which has impact either on the supply side (e.g., on venture capital firms) or on the demand side (e.g., on start-ups) or by (ii) direct public investment schemes.[20] While regulation-de-regulation aimed at creating an enabling environment for private actors to develop their activities, the direct intervention in form of investment schemes effectively enables governmental agencies to fund start-ups in particular sectors and influence investor's behavior.

Yet, clearly more is needed than investment promotion. There is now a much greater awareness of how the success of Silicon Valley is more than just about investment. There is an enormous literature that aims to provide a better understanding of what is needed, in particular seeking to identify the "ingredients" of a successful ecosystem such as Silicon Valley. The aim of developing this understanding is to recreate such an environment elsewhere.

Take Victor Hwang and Greg Horowitt's metaphor of the "rainforest."[21] In identifying the factors necessary to replicate Silicon Valley, Hwang and Horowitt emphasize the importance of a culture in which uncontrolled interactions routinely occur between talent, capital, ideas, and opportunities, i.e., the essential elements in any successful innovation ecosystem. In this type of account, innovation is an unplanned and spontaneous event a feature of the ecology of a rainforest – that is contrasted with the "planned production" of an industrial economy.

Or, Brad Feld. His book, Start-up Communities outlines what he calls the "Boulder Thesis," the elements that he feels have been key to the success of Boulder's start-up ecosystem.[22] Most importantly this means that being led by entrepreneurs. To be successful, an ecosystem must be led by the entrepreneurs themselves, not other players such as governments, universities, or investors. A second factor is long-term commitment: ecosystem builders in the community should take a long-term view, in the order of 20 years or more. Finally, Feld points to a "philosophy of inclusiveness": the ecosystem must be open and welcoming of all. The ecosystem should have regular activities that engage both new and experienced entrepreneurs, as well as investors, mentors, and more.

---

[18] See Fenwick and Vermeulen (2015b).
[19] See, e.g., Hwang and Horowitt (2012).
[20] See, e.g., Lerner (2012).
[21] Hwang and Horowitt (2012), p. 10.
[22] Feld (2016).

Or, Steve Case, Co-founder of AOL and author of The Third Wave: An Entrepreneur's Vision of the Future. He has been behind the platform celebrating and investing in emerging start-up ecosystems, the Rise of the Rest movement. In the Rise of the Rest 2018 Ecosystem Playbook, he refers to The Seven Spokes of a Start-up "Hub" seven entities that help to fuel the rise of start-up ecosystems: Local government, Universities, Investors, Start-up support organizations, Corporations, Local media, Start-ups themselves.[23]

But in order to develop a better understanding of what's needed it is also important not to generalize the issue (i.e., to focus on replicating Silicon Valley in some general sense), but also to adopt a perspective that involves looking at how ecosystems might be developed in the context of specific industries or sectors of the economy. Here we would like to suggest that, in the context of AI, a strategy building the ecosystem around established corporations might be particularly effective. In particular, we would like to suggest that whilst regulators have often focused on strategies that aim to develop ecosystems by promoting investment, they tend to neglect the role of partnerships between startups and incumbent, established corporates. Moreover, it is argued that such partnerships are particularly important in the context of "blue sky fields" such as AI.

## *4.2 Building the "Right Kind" of AI Ecosystem I: The EU Experience*

The question that therefore needs to be asked is: How to Build the "Right Kind" of AI Ecosystem? National and local governments all now see start-up ecosystems as a necessity for preparing for the future. Putting in place the necessary infrastructure to stimulate the creation, growth and scaling of new and innovative business is now seen as an important and legitimate policy objective for all levels of government.

So, what can governments do to build an effective innovation ecosystem? Traditionally, the focus of policymakers looking to create an innovation ecosystem has been on making more risk capital available for start-up and scale-up companies.

Take Europe as an example. Recently, we can see a steady decline in venture capital activity at all the stages of a start-up's development.[24] This tracks the worldwide trend (according to the data provider and analyst Pitchbook). What appears to be even more worrying is that right now the activity in first-time venture investments is the lowest for over seven years.

In order to stimulate venture capital investments, governments have used several strategies. First, there is long-standing evidence that government support has played a vital role in encouraging entrepreneurship and the launch of start-up companies. For example, governments, in their efforts to establish a sustainable ecosystem, have become the main "post-financial crisis" investor in Europe's start-up scene.

---

[23] Case (2017); for a similar argument, see Fenwick and Vermeulen (2016).
[24] See Vermeulen (2018).

Second, governments often introduce schemes that aim to activate private investments. A recent European example is a joint initiative by the European Commission and the European Investment Fund to set up a Pan-European "VC Fund-of-Funds." The investment of 25% of the total fund-size must encourage private investors, particularly, institutional investors, to invest in the next generation of innovative companies.

Third, regulatory measures can be implemented to make venture capital venture capital more accessible to investors. The proposed amendments to the European Venture Capital Fund (EU VECA) and the European Social Entrepreneurship Funds (EU SEF) are intended to give a boost to the venture capital industry in Europe. Also, investors are encouraged to make venture capital investments through fiscal incentives and tax breaks.

But even if these measures significantly increase the amount of venture capital available, entrepreneurs are not always better off. As is the case with any industry that enjoys a boom, non-specialists will emerge looking to get a piece of the growing pie. There are multiple examples of "new" venture capital investors that have started to invest in innovative companies without doing their proper homework or understanding the rules of the game.

The fear of giving up equity and losing ownership and, eventually, control to less than stellar venture capital investors only feeds a growing skepticism among entrepreneurs about attractive venture capital or other sources of risk-capital. This means that entrepreneurs prefer bootstrapping, perhaps supplemented with government grants or private loans from family, friends and fools (the "3 Fs").

Although in some cases this can be an effective model, it undoubtedly exposes founder-entrepreneurs to a much greater degree of financial risk and uncertainty. Grants can fill this investment gap but drafting and submitting proposals can take a long time. There is always a lot of competition and managing and administrating the grants can be cumbersome, costly and, ultimately a time-consuming distraction.

As such, venture capital may not represent the missing ingredient for most innovation ecosystems today. The more businesses that are created, the more money becomes available for innovation and innovative firms. There is often an extensive infrastructure supporting entrepreneurs in starting a new business.

Take the example of artificial intelligence. Most of the recently acquired AI companies come from *outside* Silicon Valley (see Figure 2). Moreover, according to CB Insights, "only" 46% of the acquired AI companies had attracted and received venture capital. So, if it is not really a question of venture capital, what is Silicon Valley doing that is missing in Europe? And what can we do to ensure the success of a sector-focused innovation ecosystem, such as an AI-oriented community of innovators?

### *4.3 Building the "Right Kind" of AI Ecosystem II: The Role of "Incumbents"*

A big part of the solution involves tapping into the experience and know-how of established enterprises. For instance, large, global established corporations often recognize that they must engage with AI, robotics and automation. They have the motive and resources to play a crucial role. And yet, all too often, existing corporate culture and governance structures mean that older, established firms struggle to adjust to new realities. Twentieth-century companies rely too heavily on hierarchical, formal and closed organizations. As such, they are ill-prepared to make the bold and agile business decisions necessary to succeed in a world of constant disruptive innovation.

To survive it is, therefore, imperative for established firms to re-invent their own innovation strategies. This means understanding how to organize for innovation, building and improving on the valuable lessons from the Silicon Valley experience. Crucially, younger firms in the innovation sector are typically organized around the kind of governance principles that provide them with the energy and ideas to constantly innovate, namely a "flat" organization, "open communication" and a "best-idea-wins-culture."[25]

Since these governance principles are more likely to be found in the organization of start-up companies, the "smartest" large corporations try to gain access to this – what Elon Musk has termed the "Silicon Valley operating system" – by cultivating open and inclusive partnerships with entrepreneurs, founders and start-ups in the innovation space. When multiple established corporations build relationships of this kind, the basis for a flourishing ecosystem can be put in place. But to build a network, community or ecosystem around this new type of partnership, two processes need to be better understood and embraced. In this way, large corporations can become the crucial link in building innovation ecosystems.

Most obviously, such linkage can drive the kind of genuine opportunities for serendipity highlighted by Hwang and Horrowit discussed above. The first strategy is for established enterprises to use corporate incubator and accelerator programs to put in place an open architecture that offers opportunities for mutual learning. Start-up founder and employees can then get to routinely mix with corporate employees. Such programs have become popular in recent years. In 2017, Amazon, Apple, Facebook, General Electric and Telefónica all announced the opening of new accelerator programs in France, India, the United Kingdom and the United States (see Figure 6).

The initial and mutual advantages of such an approach seem obvious:

i. *For Established Corporations*. Corporate incubator and accelerator programs allow large firms to engage alongside start-ups and their founders. Such collaborations give them access to ideas and strategies they would never be able to nurture internally.
ii. *For Start-up Companies*. Corporate incubator and accelerator programs are particularly interesting if there is a good cultural match. They can provide start-ups with the necessary capital and deliver tremendous resources in the form of access to relevant knowledge and established international distribution channels.

---

[25] See Fenwick and Vermeulen (2015a).

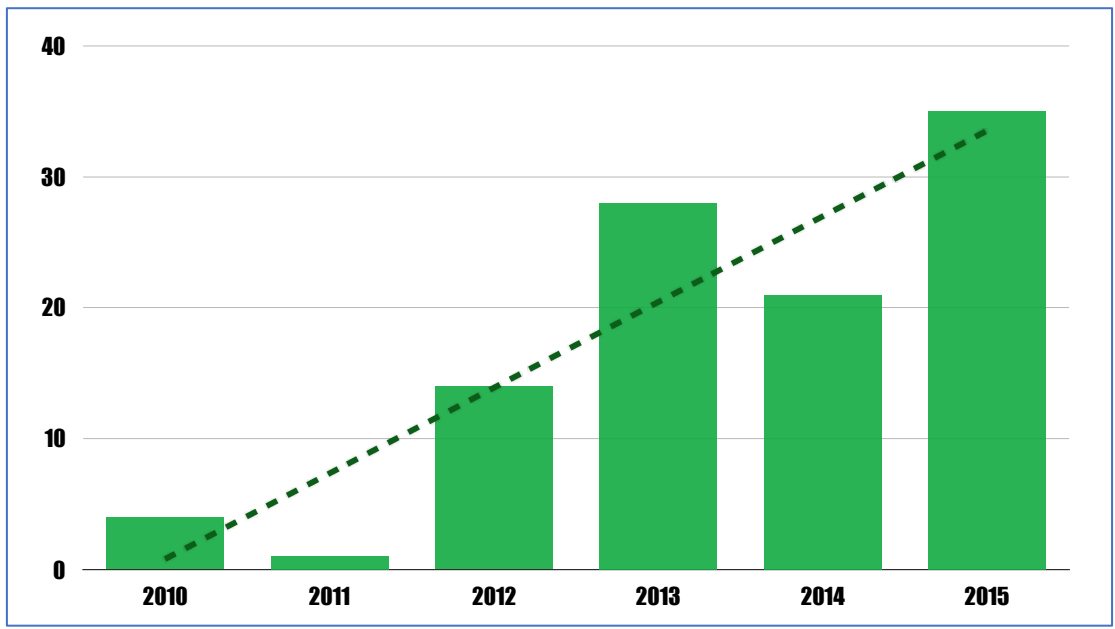

***Figure 6:*** *Launch of Corporate Incubators/Accelerators (2010-15)*[26]

Nevertheless, the success in any given case will always depend on the structure of the specific program. For instance, there are programs powered by external incubator-accelerator service providers, such as TechStars and Plug and Play. But the clear majority of programs are directly set up by the corporate hosts themselves (see Figure 4). This would indicate that innovation is more than just the process of involving start-ups.

As Moore's Law begins to slow down, clever high-tech companies are putting more emphasis on the longer-term strategies. In this respect, innovation is not, and never can be, a department. It is a culture that needs to permeate the entire enterprise. This means accepting that innovation cannot be ordered as a product. Without the right environment of clever, motivated, collaborative people, the best ideas will wither and die. Building and then sharing a vision of the domain is crucial. By putting actors together in the right way, the boundaries between corporate and start-up can be blurred, creating new opportunities for positive encounters and interaction.

The second set of strategies essential for building a successful AI ecosystem is to keep things simple and transparent. There can be no misalignment between the interests of the start-ups within the ecosystem and the interests of the corporation.

---

[26] Source: Corporate Accelerator DB/TechCrunch.

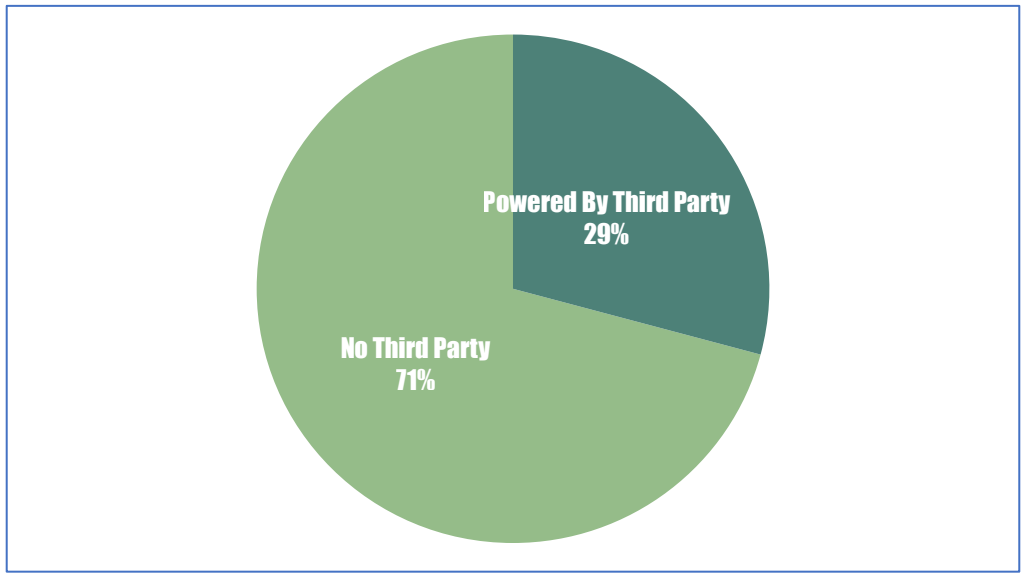

*Figure 7:* *Structure of Corporate Incubators/Accelerators (2010-16)[27]*

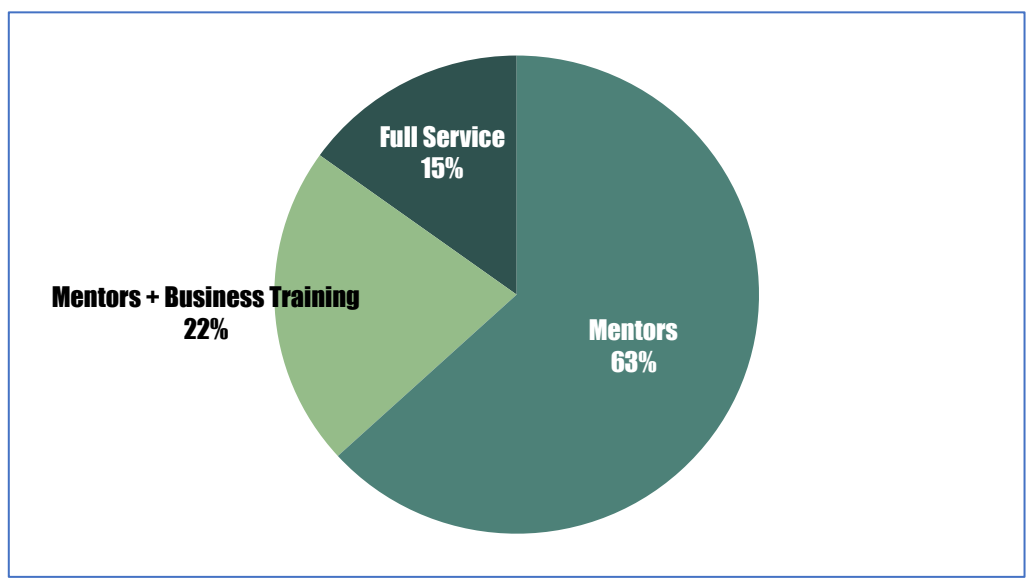

*Figure 8:* *Support Offered by Corporate Incubators/Accelerators (2010-16)[28]*

Once a startup has been accepted into the program, the main focus of the corporation should be on assisting and accelerating startups. That means actively connecting them to potentially interesting networks, customers, etc. The possible strategic returns for the corporate partners are a secondary effect or by-product of collaborating with the startups.

Moreover, by working "cheek-by-jowl" with startup founders, the corporate employees are better able to identify "out-of-the-box" solutions to specific business challenges. The potential benefits are again particularly high in a "blue sky" field such as AI where the technology is complex and "solutions" are not immediately obvious.

---

[27] Source: Corporate Accelerator DB/TechCrunch.
[28] Source: Corporate Accelerator DB/TechCrunch.

From this perspective, non-equity programs seem preferable. Avoiding ownership minority stakes can help simplify relationships, especially at a stage where neither party has a real insight into the true market value of the startup (a common phenomenon in a "blue sky" context). A corporate program that has a first (or even only) objective of making money makes the mistake of trying to execute a business model before the startup has verified that there is one. Such an approach also provides for an important ingredient in developing mutual trust. The absence of direct financial interest in participating startups can help to convince founders that the corporation will not try to appropriate their technology or limit their future options for external financing and strategic partnerships with other corporations (see Figure 6).

For instance, Microsoft accelerator does not take an equity stake in participating startups and at the same time does not require applicants' products or technology to directly complement or fit with products of Microsoft.[29] To a certain extent, Microsoft relies on serendipity occurrences in the development and future use of technology and cooperation, which may not be always foreseen or obvious.

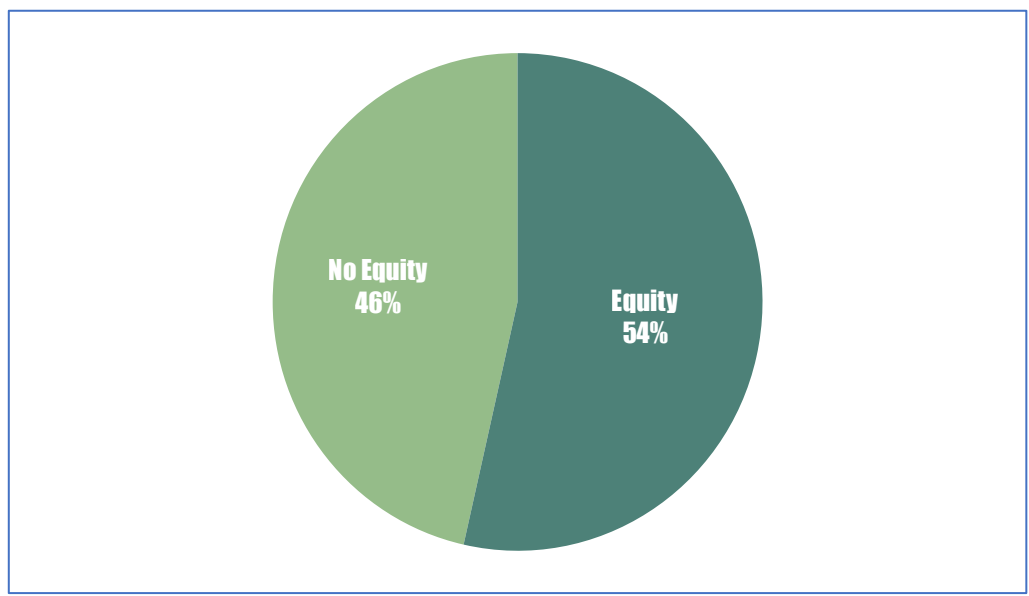

***Figure 9:*** *Equity versus Non-Equity Based Corporate Incubators (2010-16)*[30]

Managing minority stakes in portfolio startup companies is often daunting from a legal and accounting point of view. Many startups fear that accepting direct or indirect investments from a corporation will restrict their future funding opportunities and bring about the risk of "negative signaling" should the corporation decide not to support or continue the investment in the future. In this respect, it is important that the corporation sends a strong signal to the ecosystem that they are a trusted partner for startups and that they won't sacrifice a founder-entrepreneur for their own strategic or short-term financial benefits.

---

[29] Interview with Maya Grossman, Head of Global Communications, Microsoft Accelerator (Tilburg-Tel Aviv, December 7, 2016).
[30] Source: Corporate Accelerator DB/TechCrunch.

The most active corporations in the technology sector already understand their role very well. Even after they have acquired a startup, they seek to preserve that startup's unique identity (often by retaining the founders on CEO positions) and do not seek to assimilate it (which has been the conventional wisdom in M&A practice until recently). Moreover, it is precisely this kind of open and inclusive partnering that needs to be at the center of an innovation ecosystem, if it is to be effective. Although large corporations play an increasingly important role in startups ecosystems, policymakers have often been unwilling to recognize this fact.

Does that mean that policymakers are looking in the wrong direction? Not necessarily. But their view does not always offer the full picture. In the context of startups, intensive governmental support has often focused mainly on developing financial markets and catalyzing the venture capital industry. This, oftentimes, prevents regulators from identifying incumbent and successful corporates as crucial "ingredient" of all startup ecosystems.

Only recently several initiatives emerged that aim to support closer startup-corporate cooperation. For instance, the government-supported COSTA (Corporates and start-ups) initiative in the Netherlands, that attracted such corporate giants as Philips, KLM, Unilever and AkzoNobel, promotes a more intensive alliance between corporates and start-ups.[31]

While such initiatives can highlight the importance of startup-corporate cooperation, there is indeed much greater space for policymakers to introduce more measures and thus strengthen the so-called triple-helix cooperation.

## 5 Conclusion

A well-run innovation ecosystem provides multiple benefits for society in general. It creates the necessary links between complementary sources of risk finance and entrepreneurs. Also, it helps build the capacity of entrepreneurs to identify future partners that are best suited to deliver a meaningful, long-term relationship and give a young firm the best chance of developing its product and scaling successfully. Finally, such an ecosystem helps policymakers develop the "know-how" to implement more dynamic and responsive forms of regulation. In the "right kind" of ecosystem environment, flexible and inclusive processes benefit startups and established companies, regulators, experts and the public.

The ultimate goal is to prepare local and regional ecosystems for a world with a very different level of automation and artificial intelligence. Having a working strategy gives them a credible chance of competing with Silicon Valley. This chapter has identified two promising elements of such an ecosystem that are particularly relevant in the context of AI-related technologies, namely regulatory sandboxes and partnerships between AI startups and incumbent corporations. Such an approach seems to bring clear and tangible benefits to all major stakeholders in such innovation ecosystems and has the potential to deliver significant benefits to the community at large.

---

[31] Costa Program (2016).